\pdfoutput=1 \pdfsuppresswarningpagegroup=1 
\documentclass[
 aps,%
 prr,%
 reprint, twocolumn,%
 groupedaddress,%
 superscriptaddress,%
 amsfonts,%
 letterpaper,%
 footinbib,%
 floatfix,%
 longbibliography
]{revtex4-2}

\RequirePackage{amsmath,amssymb,bm}
\RequirePackage{graphicx}
\RequirePackage[usenames,dvipsnames]{xcolor}

\usepackage{pdfpages}
\makeatletter\AtBeginDocument{\let\LS@rot\@undefined}\makeatother

\usepackage[T1]{fontenc}

\usepackage{enumitem}

\usepackage{verbatim}

\usepackage{hyperref}
\hypersetup{
    colorlinks=true,
    citecolor= blue,
    linkcolor= magenta,
    filecolor=cyan,      
    urlcolor= blue,
    pdfauthor = {Chuan Chen},
    pdfcreator = {\LaTeX\ and \flqq hyperref\frqq},
}

\usepackage[capitalise]{cleveref}

\graphicspath{ {./Figs/} }

\newcommand{\Fref}[1]{Fig.~\ref{#1}}

\newcommand{\Eqref}[1]{Eq.~\eqref{#1}}

\newcommand{\Secref}[1]{Sec.~\ref{#1}}

\newcommand{\bX}{\mathbf{X}}
\newcommand{\bZ}{\mathbf{Z}}
\newcommand{\bbZ}{\mathbb{Z}}
\newcommand{\NE}{\text{NE}}
\newcommand{\NW}{\text{NW}}
\newcommand{\SW}{\text{SW}}
\newcommand{\SE}{\text{SE}}
\newcommand{\N}{\text{N}}
\newcommand{\W}{\text{W}}
\newcommand{\E}{\text{E}}
\newcommand{\bk}{\mathbf{k}}

\makeatletter
\newcommand*{\rom}[1]{\expandafter\@slowromancap\romannumeral #1@}
\makeatother

\newcommand{\teq}{{\,=\,}}

\newcommand{\tlt}{{\,<\,}}

\begin{document}

\title{Berry Phases of Vison Transport in $\mathbb{Z}_2$ Topologically Ordered States from
Exact Fermion-Flux Lattice Dualities}

\author{Chuan~Chen}
\affiliation{
Institute for Advanced Study, Tsinghua University,
100084 Beijing, China
}
\affiliation{
Max-Planck Institute for the Physics of Complex Systems,
01187 Dresden, Germany
}

\author{Peng Rao}
\affiliation{
Max-Planck Institute for the Physics of Complex Systems,
01187 Dresden, Germany
}

\author{Inti~Sodemann}
\email{sodemann@itp.uni-leipzig.de}
\affiliation{
Max-Planck Institute for the Physics of Complex Systems,
01187 Dresden, Germany
}
\affiliation{Institut f\"ur Theoretische Physik,
Universit\"at Leipzig,
04103 Leipzig, Germany
}

\date{\today}

\begin{abstract}
We develop an exact map of all states and operators from 2D lattices of spins-$1/2$ into lattices
of fermions and bosons with mutual semionic statistical interaction that goes beyond previous
dualities of $\mathbb{Z}_2$ lattice gauge theories because it does not rely on imposing
local conservation laws and captures the motion of ``charges'' and ``fluxes'' on equal footing.
This map allows to explicitly compute the Berry phases for the transport of fluxes in 
a large class of symmetry enriched topologically ordered states with emergent $\mathbb{Z}_2$ gauge fields
that includes chiral, non-chiral, abelian or non-abelian, that can be perturbatively connected
to models where the visons are static and the emergent fermionic spinons have a non-interacting dispersion.
The numerical complexity of computing such vison phases reduces therefore to computing overlaps
of ground states of free-fermion Hamiltonians.
Among other results, we establish numerically the conditions under which the Majorana-carrying
flux excitation in Ising-Topologically-Ordered states enriched by translations acquires $0$
or $\pi$ phase when moving around a single plaquette.
\end{abstract}

\maketitle

\section{Introduction}
One of the best understood families of spin liquids are those featuring emergent
$\bbZ_2$ gauge fields~\cite{Wen2010,Fradkin2013}. These spin liquids, which include the
original Anderson short-ranged RVB state~\cite{Anderson1987,Baskaran1987}, feature a
non-local fermion (spinon) and a ``$\pi$-flux'' (vison)
excitation~\cite{Read1989,  Kivelson1989, Read1991, Senthil2000}.
Kitaev's toric code (TC)~\cite{Kitaev2003} is perhaps the simplest exactly solvable
model for these kind of spin liquids.
A recent series of works~\cite{Chen2018, Pozo2021, Rao2021} have shown that,
beyond being an exactly solvable model, the TC offers a new way to organize
the Hilbert space.
In Ref.~\cite{Chen2018}, it has been shown that by imposing a new type of local $\bbZ_2$ 
constraint (local symmetry) on a spin model, the local gauge invariant spin
operators can be exactly mapped onto local fermion bilinears.
This construction can be viewed as a generalization of the procedure that allows
to solve the Kitaev honeycomb model exactly~\cite{Kitaev2006}, where the $\bbZ_2$
constraint immobilizes the flux excitations leaving the fermions as the only dynamical
objects of the problem.
For related constructions see Refs.~\cite{Kitaev2006, Chen2007, Chen2008, Cobanera2010,
Cobanera2011, Nussinov2012}. The construction of Ref.~\cite{Chen2018} provides a local
map from fermion bilinear operators onto spin operators in 2D, and it serves to rewrite
in an exact manner any imaginable local Hamiltonian of fermions as a local Hamiltonian 
of spins restricted to a subspace satisfying the $\bbZ_2$ local conservation laws.
Thus, for example, any free fermion model can be obtain as an exact description of
a subspace of the Hamiltonian of a spin model.

In this work we extend the mapping of Ref.~\cite{Chen2018} by constructing an exact
lattice duality mapping of the \emph{full} Hilbert space of the underlying spins onto a
dual space of spinons and visons \emph{without} imposing any local $\mathbb{Z}_2$
conservation laws that would freeze the motion of these particles. 
Namely, we will construct non-local spinon and vison creation/annihilation operators
in a completely explicit form in terms of underlying spin-$1/2$ operators.
One of the key properties of our construction is that the dual Hilbert space completely
``disentangles'' the vison and emergent fermion degrees of freedom, in the sense that
the dual states can be organized as tensor products of vison and emergent fermion
configurations.
We will use this construction to compute the Berry phases associated with
transporting the vison around plaquettes in closed loops in the background of
topological superconducting state of the spinons with a non-zero Chern number.
Throughout this work we will refer to the vison $\pi$-flux excitations sometimes as
``$e$-particles'' and the fermionic spinons as the ``$\varepsilon$-particles''.
A recent work~\cite{Rao2021} computed these phases when the fermions were in BdG states
with zero Chern number, relying on the property that these could be realized as ground states of
commuting projector Hamiltonians. But it is known that chiral states cannot be realized in this
fashion~\cite{Kapustin2020}, and therefore our current approach overcomes these limitations.

The rest of the paper is organized as follows:
\Secref{sec:duality} contains the theoretical foundation of this work, which is an
exact duality mapping of a 2D spin system and
a Hilbert space of mutual semions. In \Secref{sec:boson-boson mapping}, we introduce
the duality mapping where the dual space consists of $e$ (boson) and $m$ (boson) particles;
in \Secref{sec:boson-fermion mapping}, we introduce the mapping with the dual space
containing visons ($e$ boson) and spinons ($\varepsilon$ fermion).
As an application of this new theoretical tool, we computed the vison Berry phases
for the celebrated Kitaev model. The model (and its dual form) was briefly reviewed
in \Secref{sec:model}. Our main results are presented in \Secref{sec:BP-visons}:
\Secref{sec:Kitaev-Haldane} shows the results for the Kitaev model with a finite spinon Haldane mass
term, and the results for a model with a higher spinon Chern number ($C = -2$)
are presented in \Secref{sec:higher-Chern}. 
We summarize and discuss our findings in \Secref{sec:discuss}.

\section{Duality mapping} \label{sec:duality}
\subsection{Boson-boson mapping} \label{sec:boson-boson mapping}
%
\begin{figure}[t]
\centering
\includegraphics[width=0.49\textwidth]{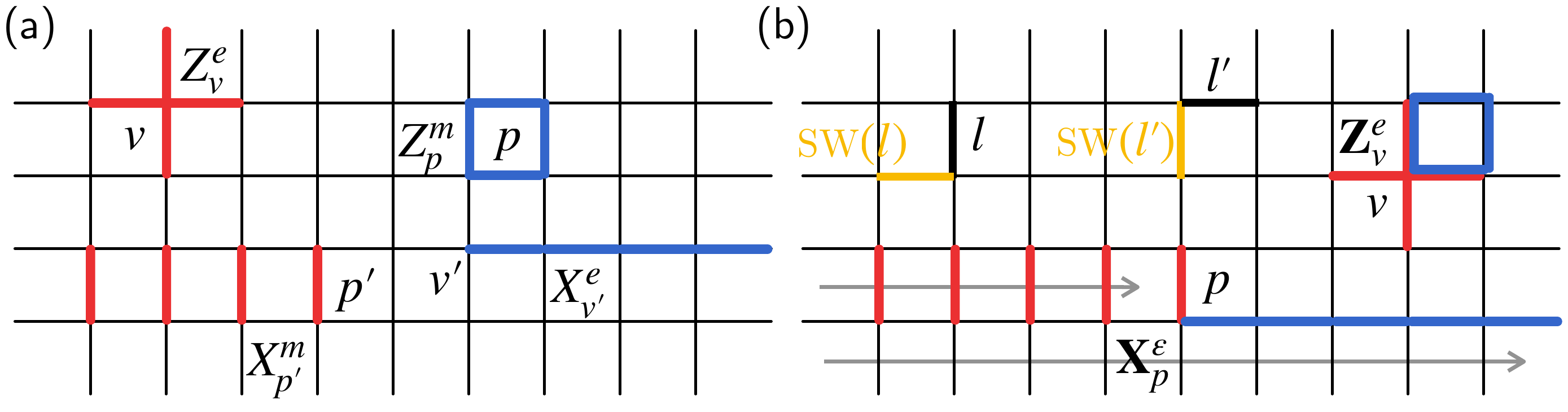}
\caption{
Schematic of the infinite system for both boson-boson
and boson-fermion mappings.
(a) Operators for the boson-boson mapping (infinite lattice).
Spin $X$ ($Z$) operators
at each link are represented by red (blue) colored bonds.
(b) Operators for the boson-fermion mapping (infinite lattice).
The gray arrow indicates the sequence of plaquettes in the Jordan-Wigner transformation,
which increases from lower to upper rows.
}
\label{fig:schematic}
\end{figure}
%
In this section, we will illustrate the idea of the boson-boson mapping.
For simplicity, here we will focus on the case of an \emph{infinite} system,
the mappings on an open and a periodic system are provided
in the Supplementary Information~\cite{SI}.

As has been shown explicitly in the TC model~\cite{Kitaev2003}, for a 2D spin system
with spins residing on the links of a square lattice (its
Hilbert space will be denoted as $\mathcal{H}_\text{spin}$), one can defined the
so-called star and plaquette operators associated with each vertex and plaquette respectively:
\begin{equation} \label{eq:star-plaquette1}
A_v = \prod_{l \in \text{star}(v)} X_l, \
B_p = \prod_{l \in \text{boundary}(p)} Z_l.
\end{equation}
All the $A_v$ and $B_p$ commute with each other, moreover,
the eigenstates of them form a basis of the spin Hilbert
space $\mathcal{H}_\text{spin}$.
One can define a dual spin system
$\mathcal{H}_{e \text{-spin}} \times \mathcal{H}_{m \text{-spin}}$
containing two types of \emph{spins}, denoted as $e$ and $m$ spins.
Here the $e$ and $m$ spins sit on the vertices and plaquettes of
the square lattice respectively, whose spin $Z$ configurations
are related to the occupation of the $e$ and $m$ particles (see below).
The local spin $X$ and $Z$ Pauli matrices of the dual $e$ ($m$) spins are denoted
as $X_v^e$ ($X_p^m$) and $Z_v^e$ ($Z_p^m$), which satisfy the following
commutation relations:
\begin{subequations}\label{eq:e-m-commute1}
\begin{align}
& [ X_v^e, Z_{v'}^e ] = 0 \ (v \neq v'),\ \{ X_v^e, Z_v^e \} = 0, \\
& [ Z_p^m, X_{p'}^m ] = 0 \ (p \neq p'),\ \{ Z_p^m, X_p^m \} = 0,\\
& [ X_v^e, X_p^m ] = [ Z_v^e, Z_p^m ] = [ X_v^e, Z_p^m ] = [ Z_v^e, X_p^m ] = 0. 
\end{align}
\end{subequations}
Within the duality mapping, we shall map the eigenbasis of $A_v$ and $B_p$
from $\mathcal{H}_\text{spin}$ to the local spin $Z$ eigenbasis of 
$e$ and $m$ spins, such that the star and plaquette operators are mapped to
the spin $Z$ Pauli matrices of $e$ and $m$ spins respectively:
\begin{equation} \label{eq:AvBp map}
A_v \leftrightarrow Z_v^e, \
B_p \leftrightarrow Z_p^m.
\end{equation}
To make the dual operators of $X_v^e$, $Z_v^e$, $X_p^m$ and $Z_p^m$ should also
satisfy the algebraic relations in \Eqref{eq:e-m-commute1},
we found that the following choice of dual operators do the job:
\begin{equation}\label{eq:Ze-Xm}
\prod_{l \in R(v)} Z_l \leftrightarrow X_v^e,
\quad
\prod_{l \in L(p)} X_l \leftrightarrow X_p^m.
\end{equation}
Here $R(v)$ stands for the horizontal links to the right of vertex $v$,
$L(p)$ stands for the vertical links to the left of plaquette $p$
(see \Fref{fig:schematic}(a) for a schematic of the non-local operators
above). 
It can be shown that the local spin $X$ and $Z$ operators in
$\mathcal{H}_\text{spin}$ can be mapped to:
\begin{enumerate}[label = \roman*).]
\item Vertical $l$:
\begin{align}
X_l \leftrightarrow X_{p_1}^m X_{p_2}^m, \quad
Z_l \leftrightarrow X_{v_1}^e X_{v_2}^e \prod_{p \in R(l)} Z_p^m.
\end{align}
\item Horizontal $l$:
\begin{align}
X_{l} \leftrightarrow X_{p_1}^m X_{p_2}^m \prod_{v \in L(l)} Z_v^e, \quad
Z_{l} \leftrightarrow X_{v_1}^e X_{v_2}^e.
\end{align}
\end{enumerate}
Here for any vertical (horizontal) link $l$, the two vertices connected by it are
denoted as $v_1$ and $v_2$, the plaquettes to its left (top) and right (bottom)
are called $p_1$ and $p_2$. $L(l)$ stands for vertices to the left of a horizontal
$l$ (including $v_1$).
Note that spin $X_l$ and $Z_l$ operators form a complete algebraic basis
out of which any other spin operators can be written in terms of their
summation, products and multiplication with complex numbers.
In this way, we have established the duality mapping between
$\mathcal{H}_{\text{spin}}$ and $\mathcal{H}_{e \text{-spin}}
\times \mathcal{H}_{m \text{-spin}}$.

Since spin-$1/2$ degrees of freedom can be equivalently viewed as 
hard-core ($e$ and $m$) bosons, it is straightforward to establish
the mapping
$\mathcal{H}_{e\text{-spin}} \times \mathcal{H}_{m \text{-spin}}
\leftrightarrow \mathcal{H}_e \times \mathcal{H}_m$,
the $e$ and $m$ spins' Pauli matrices can be written as
bosonic operators:
\begin{subequations}\label{eq:map-em1}
\begin{align}
& Z_v^e \leftrightarrow (-1)^{b_v^\dagger b_v},
\ X_v^e \leftrightarrow (b_v + b_v^\dagger), \\
& Z_p^m \leftrightarrow (-1)^{d_p^\dagger d_p},
\ X_p^m \leftrightarrow (d_p + d_p^\dagger).
\end{align}
\end{subequations}
Here $b_v$ ($d_p$) is the annihilation operator of an $e$ ($m$) boson
at vertex $v$ (plaquette $p$).

Finally, we obtain the duality mapping between $\mathcal{H}_{\text{spin}}$
and $\mathcal{H}_e \times \mathcal{H}_m$, where the star and plaquettes operators
of the original spin space are mapped to the parity operators of $e$ and $m$ bosons:
\begin{equation} \label{eq:AvBp-boson}
A_v \leftrightarrow (-1)^{b_v^\dagger b_v}, \
B_p \leftrightarrow (-1)^{d_p^\dagger d_p}.
\end{equation}
The local spin operators are mapped into:
\begin{enumerate}[label = \roman*).]
\item Vertical $l$:
\begin{align}
& X_l \leftrightarrow (d_{p_1} + d_{p_1}^\dagger) (d_{p_2} + d_{p_2}^\dagger), \\
& Z_l \leftrightarrow (b_{v_1} + b_{v_1}^\dagger) (b_{v_2} + b_{v_2}^\dagger)
\prod_{p \in R(l)} (-1)^{d_p^\dagger d_p}.
\end{align}
\item Horizontal $l$:
\begin{align}
& X_l \leftrightarrow \prod_{v \in L(l)} (-1)^{b_v^\dagger b_v} \ 
(d_{p_1} + d_{p_1}^\dagger) (d_{p_2} + d_{p_2}^\dagger), \\
& Z_l \leftrightarrow (b_{v_1} + b_{v_1}^\dagger) (b_{v_2} + b_{v_2}^\dagger).
\end{align}
\end{enumerate}
Within this duality mapping, local $Z_l$ ($X_l$) operators have the effect
of pair fluctuating and hopping the $e$ ($m$) particles on nearest-neighbor
vertices $v_1$ and $v_2$ (plaquettes $p_1$ and $p_2$), as one would naturally
expect since they anti-commute with the $e$ ($m$) particles' parity operator at those
two vertices (plaquettes). More interestingly, there is also a product of $m$ ($e$)
particles' parity operators when the $e$ ($m$) particle is hopping along
$y$-direction~\footnote{Here the fact that the statistical interaction only
appears when $e$ and $m$ particles are pair fluctuating or hopping along the
$y$-direction can be understood as due to the specific ``gauge choice'' in our duality
mapping: each $e$ ($m$) particle produces
a ``branch-cut'' in the vector potential to the right (left) infinity, which
is felt by the $m$ ($e$) particles.},
such non-local statistical interaction terms make the $e$ and $m$
particles mutual semions.

\subsection{Boson-fermion mapping} \label{sec:boson-fermion mapping}
\subsubsection{Infinite lattice}
It turns out that it is also possible to map the $\mathcal{H}_\text{spin}$
to a space of bosons ($e$ particles) and fermions ($\varepsilon$ particles),
$\mathcal{H}_e \times \mathcal{H}_\varepsilon$. 
For pedagogical reason, we will start with the case with an infinite
lattice, and introduce the mapping on a periodic system in the next section.
The mapping for an open lattice can be found in the Supplementary
information~\cite{SI}.

Each $\varepsilon$ particle in the boson-fermion mapping can be viewed as
a composite of an $e$ and an $m$ particles of the boson-boson mapping,
and the $e$ and $\varepsilon$ particles are mutual semions~\cite{Kitaev2003}.
Same as the boson-boson mapping introduced in the previous section,
the mapping between $\mathcal{H}_\text{spin}$ and
$\mathcal{H}_e \times \mathcal{H}_\varepsilon$
can be made more obvious if one first introduces an intermediate dual spin space
$\mathcal{H}_{e\text{-spin}} \times \mathcal{H}_{\varepsilon \text{-spin}}$,
where the $e$ ($\varepsilon$) spins are located at the vertices
(plaquettes) of a square lattice.
Recall that the eigenstates of $A_v$ and $B_p$ are also eigenstates of
all the $A_v B_{\mathsf{p}(v)}$ and $B_p$. Here $\mathsf{p}(v)$ stands for the
plaquettes to the northeast of vertex $v$.
One can map this eigenbasis to the local spin $Z$ eigenbasis of $e$ and
$\varepsilon$ spins of the intermediate dual space, such that
\begin{equation}
A_v B_{\mathsf{p}(v)} \leftrightarrow \bZ_v^e, \
B_p \leftrightarrow \bZ_p^\varepsilon.
\end{equation}
Note that here we used bold symbols to denote the Pauli
matrices of the $e$ and $\varepsilon$ spins, which satisfy the
following algebraic relations:
\begin{subequations} \label{eq:eep-XZ-algebra}
\begin{align}
& [ \bX_v^e, \bZ_{v'}^e ] = 0 \ (v \neq v'), \
\{ \bX_v^e, \bZ_v^e \} = 0, \\
& [ \bZ_p^\varepsilon, \bX_{p'}^\varepsilon ] = 0 \ (p \neq p'), \
\{ \bZ_p^\varepsilon, \bX_p^\varepsilon \} = 0, \\
& [ \bZ_v^e, \bZ_p^\varepsilon ] = 
[ \bX_v^e, \bX_p^\varepsilon ] = 
[ \bX_v^e, \bZ_p^\varepsilon ] =
[ \bZ_v^e, \bX_p^\varepsilon ] = 0.
\end{align}
\end{subequations}
In $\mathcal{H}_{\text{spin}}$, the dual operators of
$\bX_v^e$ and $\bX_p^\varepsilon$ will respect these relations
if one choose:
\begin{align}
& \prod_{l \in R(v)} Z_l \leftrightarrow \bX_v^e, \\
& \prod_{l \in R(\mathsf{v}(p))} Z_l \prod_{l' \in L(p)} X_{l'}
\leftrightarrow \bX_p^\varepsilon.
\end{align}
Here $\mathsf{v}(p)$ stands for the vertex to the southwest of plaquette $p$.
A schematic of these non-local spin operators are shown in \Fref{fig:schematic}(b).
In this way, we have completed the mapping between $\mathcal{H}_\text{spin}$
and $\mathcal{H}_{e\text{-spin}} \times \mathcal{H}_{\varepsilon \text{-spin}}$.

The mapping from $\mathcal{H}_{e\text{-spin}} \times \mathcal{H}_{\varepsilon \text{-spin}}$
to $\mathcal{H}_{e} \times \mathcal{H}_{\varepsilon}$ is more straightforward,
the $e$ particles is just the hard-core boson corresponding to the $e$ spins, and the
$\mathcal{H}_{\varepsilon \text{-spin}}$ is mapped to $\mathcal{H}_\varepsilon$
through a Jordan-Wigner transformation:
\begin{align} 
& \bZ_v^e \leftrightarrow (-1)^{b_v^\dagger b_v},  \ 
\bX_v^e \leftrightarrow (b_v + b_v^\dagger), \label{eq:esp-e} \\
& \bZ_p^\varepsilon \leftrightarrow -i \gamma_p \gamma'_{p}, \ 
\bX_p^\varepsilon \leftrightarrow \left( \prod_{q < p} -i \gamma_q \gamma'_q \right) \gamma'_p \label{eq:epsp-ep}.
\end{align}
Here $b_v$ ($b_v^\dagger$) is the annihilation (creation) operator for the $e$ paticle
at vertex $v$. We have also introduced two Majorana fermion modes ($\gamma_p$ and $\gamma'_p$)
to represent the complex $\varepsilon$ fermion mode
(whose annihilation/creation operator is $c_p/c_p^\dagger$) at each plaquette $p$, with
\begin{equation}
\gamma_p = c_p + c_p^\dagger, \ \gamma'_p = \frac{1}{i}( c_p - c_p^\dagger ).
\end{equation}
Note that the fermion partiy at each plaquette $p$ is
$(-1)^{c_p^\dagger c_p} = -i \gamma_p \gamma'_p$.
The sequence of plaquettes in the Jordan-Wigner transformation is indicated by the gray arrow
in \Fref{fig:schematic}(b).
In this way, we have established the mapping between $\mathcal{H}_\text{spin}$ and
$\mathcal{H}_e \times \mathcal{H}_\varepsilon$, it can be shown that the following local
spin operators are mapped to:
\begin{enumerate}[label = \roman*).]
\item $l$ is a vertical link:
\begin{align}
&X_l Z_{\SW(l)} \leftrightarrow \bX_{p_1}^\varepsilon \bX_{p_2}^\varepsilon
\leftrightarrow i \gamma_{p_1} \gamma'_{p_2}, \\
&Z_l \leftrightarrow \bX_{v_1}^e \bX_{v_2}^e \prod_{p \in R(l)} \bZ_p^\varepsilon \nonumber \\
& \leftrightarrow (b_{v_1} + b_{v_1}^\dagger)(b_{v_2} + b_{v_2}^\dagger)
\prod_{p \in R(l)} ( -i\gamma_p \gamma'_p ).
\end{align}
\item $l$ is a horizontal link:
\begin{align}
& X_{l} Z_{\SW(l)} \leftrightarrow (-1) \prod_{v \in L(l)} \bZ_v^e
\left( \prod_{p_2 \le p \le p_1} \bZ_p^\varepsilon \right) \
\bX_{p_1}^\varepsilon \bX_{p_2}^\varepsilon
\nonumber \\
& \leftrightarrow \prod_{v \in L(l)} (-1)^{b_v^\dagger b_v} \ i\gamma_{p_1} \gamma'_{p_2}
\\
& Z_{l} \leftrightarrow \bX_{v_1}^e \bX_{v_2}^e \leftrightarrow (b_{v_1} + b_{v_1}^\dagger)
(b_{v_2} + b_{v_2}^\dagger).\label{eq:bf-mapping-Z_l}
\end{align}
\end{enumerate}
Here $\SW(l)$ is the link to the southwest of link $l$, which also connects to it (see
\Fref{fig:schematic}(b) for a schematic). It is clear that the local $Z_l$ ($X_l Z_\SW(l)$)
operator is able to pair create, annihilate and hop the $e$ ($\varepsilon$) particles
in the nearest neighbors. The non-local products of the $e$-particle ($\varepsilon$-particle)
parities in the dual operator of $X_l Z_{\SW(l)}$ ($Z_l$) indicates the statistical
interaction between between $e$ and $\varepsilon$ particles, which view each other
as $\pi$ fluxes, i.e., they are mutaul semions.

\subsubsection{Periodic lattice}
%
\begin{figure}[t]
\centering
\includegraphics[width=0.49\textwidth]{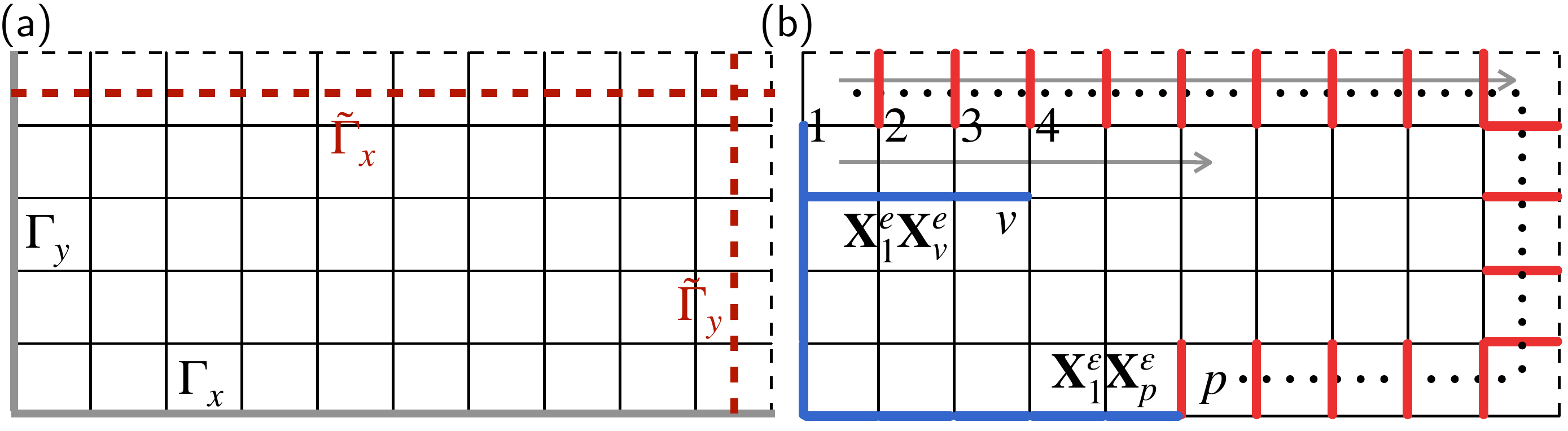}
\caption{
Schematic of operators in the boson-fermion mapping for a periodic lattice.
(a) Non-contractible loops used in the definition of Wilson loop
and t'Hooft operators. $\Gamma_{x/y}$ is highlighted in gray color and
$\tilde{\Gamma}_{x/y}$ is highlighted in red color.
(b) The dual of $\bX_{1}^e \bX_{v}^e$ and $\bX_{1}^\varepsilon \bX_{p}^\varepsilon$
operators. The sequence of vertices and the associated plaquettes (to the northeast of
each vertex) starts with the one on the top left, ascends towards the right direction
within each row and increases from the top to bottom rows, as indicated by the gray arrow.
The $\Gamma_{1,v}$ path starts from vertex $1$, goes down first then goes to the right
direction until reaching vertex $v$, see the path paved by blue coloured bonds.
The dual path $\tilde{\Gamma}_{1,p}$ starts from plaquette $1$, goes to the
right end first, then goes down, and finally goes left until plaquette $p$.
An example is indicated by the black dotted line in the figure.
}
\label{fig:schematic-torus}
\end{figure}
%
The idea of the duality mapping on a periodic lattice (torus) is basically the same
as the infinite lattice case. However, there are now two global constraints in the original spin
space:
\begin{equation} \label{eq:constraint-eepsilon}
\prod_v A_v B_{\mathsf{p}(v)} = 1, \ 
\prod_p B_p = 1.
\end{equation}
Therefore, only an even number of $A_v B_{\mathsf{p}(v)}$ and $B_p$
can take $-1$, i.e., there are only $2^{2 \mathcal{N} - 2}$ 
different configurations of $A_v B_{\mathsf{p}(v)}$
and $B_p$, where $\mathcal{N}$ is the number of unit cells in the system.
To fully characterize the spin Hilbert space (with dimension $2^{2 \mathcal{N}}$),
one needs to introduce two additional Wilson loop degrees of freedom.
The Wilson loop operators commutes with all the $A_v B_{\mathsf{p}(v)}$ and
$B_p$ operators, one possible choice is:
\begin{align}
W_1 = - \prod_{l \times \tilde{\Gamma}_x} X_l
\prod_{l' \in \Gamma_x} Z_{l'}, \ 
W_2 = - \prod_{l \times \tilde{\Gamma}_y} X_l
\prod_{l' \in \Gamma_y} Z_{l'}.
\end{align}
Here $l \times \tilde{\Gamma}_{x/y}$ denotes the link $l$ crossing the dual-lattice
path $\tilde{\Gamma}_{x/y}$.
Paths $\Gamma_{x,y}$ and $\tilde{\Gamma}_{x,y}$ are shown in
\Fref{fig:schematic-torus}(a).
$W_{1/2}$ takes the value of $\pm 1$ and can be interpreted as a closed
transport of $\varepsilon$-particles across a $x/y$-oriented non-contractible
loop of the torus (see below).
One can also define two t'Hooft operators $T_1$ and $T_2$ which commutes with 
all the $A_v B_{\mathsf{p}(v)}$ and $B_p$ but respectively anti-commutes with
$W_1$ and $W_2$, which read:
\begin{equation}
T_1 = \prod_{l \in \Gamma_y} Z_l, \quad T_2 = \prod_{l \in \Gamma_x} Z_l.
\end{equation}
As will become clear later,
$T_{1/2}$ plays the role of transporting an $e$-particle across the $y/x$-oriented
non-contractible loop of the torus.

The intermediate dual (spin) space for a periodic system reads
$\mathcal{H}_{e \text{-spin}}^{\text{even }\downarrow} \times
\mathcal{H}_{\varepsilon \text{-spin}}^{\text{even }\downarrow} \times
\mathcal{H}_W$.
Here $\mathcal{H}_{e \text{-spin}}^{\text{even } \downarrow}$ stands for the even-$\downarrow$
subspace of the $e$ spins (same for the
$\mathcal{H}_{\varepsilon \text{-spin}}^{\text{even } \downarrow}$) due to the constraint
\Eqref{eq:constraint-eepsilon}.
$\mathcal{H}_W$ is a 4-dimension Hilbert space containing two (auxiliary) spins, which we
call Wilson loop spins (WLS) as they corresponds to the two Wilson loop degrees of freedom
in the original spin system.
When establishing the mapping between $\mathcal{H}_{\text{spin}}$ and
$\mathcal{H}_{e \text{-spin}}^{\text{even }\downarrow} \times
\mathcal{H}_{\varepsilon \text{-spin}}^{\text{even }\downarrow} \times
\mathcal{H}_W$, the eigenbasis of all the $A_v B_{\mathsf{p}(v)}$,
$B_p$ and $W_{1,2}$ will be mapped to the spin $Z$ eigenbasis
of $e$ spins, $\varepsilon$ spins and WLS, which gives:
\begin{equation}
A_v B_{\mathsf{p}(v)} \leftrightarrow \bZ_v^e,\ 
B_p \leftrightarrow \bZ_p^\varepsilon, \ 
W_{1,2} \leftrightarrow Z_{1,2}^W.
\end{equation}
The t'Hooft operators are mapped to the Pauli $X$ matrices of the WLS:
$T_{1,2} \leftrightarrow X_{1,2}^W$.
Note that there is an implicit projection operator $P$ in the dual spin
operators, which projects states to the even-$\downarrow$ subspace of $e$ and $\varepsilon$
spins.
Since the physical dual spin subspace states contains only an even number of
($e$ and $\varepsilon$) down spins, a single $\bX_v^e$ or $\bX_p^\varepsilon$
has no matrix element in this subspace because they only mix states
with different number of down spins.
On the other hand, bilinears of $\bX_v^e$ or $\bX_p^\varepsilon$ have non-zero matrix elements
in the physical subspace.
For convenience, we take vertex/plaquette $1$ as a
``reference'' vertex/plaquette (see \Fref{fig:schematic-torus}(b)), and looked for the
dual operators of $\bX_1^e \bX_v^e$ and $\bX_1^\varepsilon \bX_p^\varepsilon$
such that the algebraic relations in \Eqref{eq:eep-XZ-algebra}
can be satisfied. One possible choice is the following mapping:
\begin{align}
& \prod_{l\in \Gamma_{1,v}} Z_l \leftrightarrow \bX_{1}^e \bX_{v}^e, \\
& \prod_{l\in \Gamma_{1,v}} Z_l
\prod_{l' \in \tilde{\Gamma}_{1,p}} X_{l'} \leftrightarrow
\bX_{1}^\varepsilon \bX_{p}^\varepsilon
\end{align}
Here $\Gamma_{1,v}$ ($\tilde{\Gamma}_{1,p}$) is a direct (dual) lattice path
connecting the vertices $1$ and $v$ (plaquettes $1$ and $p$), see
\Fref{fig:schematic-torus}(b) for a schematic of them.
To simplify the notation, we are simply using the sequence numbers of 
vertices and plaquettes to denote them in the subindices of the operators (see
their order in \Fref{fig:schematic-torus}(b)).

The mapping from $e$ spins ($\varepsilon$ spins) to the $e$ bosons ($\varepsilon$ fermions)
is very similar to the infinite lattice case shown in \cref{eq:esp-e,eq:epsp-ep}, however,
due to the constraints in \Eqref{eq:constraint-eepsilon}, the (physical) $e$- and
$\varepsilon$-particle states contains only an even number of particles.
The dual boson-fermion (and WLS) space reads: 
$\mathcal{H}_e^\text{even} \times \mathcal{H}_\varepsilon^\text{even} \times \mathcal{H}_W$.
Note that the sequence of plaquettes in the Jordan-Wigner transformation 
between $\varepsilon$ spins and $\varepsilon$ fermions has also
changed now (which is shown in \Fref{fig:schematic-torus}(b)).
In this way, one obtains the duality mapping between $\mathcal{H}_\text{spin}$ 
and $\mathcal{H}_e^\text{even} \times \mathcal{H}_\varepsilon^\text{even} 
\times \mathcal{H}_W$, local $X_l$ and $X_l Z_{\SW(l)}$ operators are mapped
to:

\noindent I). $l$ is a vertical link
\begin{enumerate}[label=\roman*).]
\item $l \notin \Gamma_y$ and $l$ does not cross $\tilde{\Gamma}_x$.
\begin{subequations}
\begin{align}
& Z_l \leftrightarrow (b_{v_1} + b_{v_1}^\dagger) (b_{v_2} + b_{v_2}^\dagger)
\left( \prod_{l \times \tilde{\Gamma}_{1,p}} -i \gamma_p \gamma'_p \right), \\
& X_l Z_{\SW(l)} \leftrightarrow i\gamma_p \gamma'_p.
\end{align}
\end{subequations}

\item $l \in \Gamma_y$ and $l$ does not cross $\tilde{\Gamma}_x$.
\begin{subequations}
\begin{align}
& Z_l \leftrightarrow (b_{v_1} + b_{v_1}^\dagger) (b_{v_2} + b_{v_2}^\dagger), \\
& X_l Z_{\SW(l)} \leftrightarrow \left[ \prod_{l \in \Gamma_{1,v}} (-1)^{b_v^\dagger b_v} \right] \
i\gamma_{p_1} \gamma'_{p_2} \ Z^W_1.
\end{align}
\end{subequations}

\item $l \notin \Gamma_y$ and $l$ crosses $\tilde{\Gamma}_x$.
\begin{subequations}
\begin{align}
& Z_l \leftrightarrow (b_{v_1} + b_{v_1}^\dagger) (b_{v_2} + b_{v_2}^\dagger) \
\left( \prod_{l \times \tilde{\Gamma}_{1,p}} -i \gamma_p\gamma'_p \right) \ X^W_1, \\
& X_l Z_{\SW(l)} \leftrightarrow i\gamma_{p_1} \gamma'_{p_2}.
\end{align}
\end{subequations}

\item $l \in \Gamma_y$ and $l$ crosses $\tilde{\Gamma}_x$.
\begin{subequations}
\begin{align}
& Z_l \leftrightarrow (b_{v_1} + b_{v_1}^\dagger) (b_{v_2} + b_{v_2}^\dagger) \ X^W_1, \\
& X_l Z_{\SW(l)} \leftrightarrow i\gamma_p \gamma'_p \ Z^W_1.
\end{align}
\end{subequations}
\end{enumerate}

\noindent II). $l$ is a horizontal link
\begin{enumerate}[label = \roman*).]
\item $l \notin \Gamma_x$ and $l$ does not cross $\tilde{\Gamma}_y$.
\begin{subequations}
\begin{align}
& Z_l \leftrightarrow  (b_{v_1} + b_{v_1}^\dagger) (b_{v_2} + b_{v_2}^\dagger), \\
& X_l Z_{\SW(l)} \leftrightarrow \left[ \prod_{l \in \Gamma_{1,v}} (-1)^{b_v^\dagger b_v} \right] \ 
i\gamma_{p_1} \gamma'_{p_2}.
\end{align}
\end{subequations}

\item $l \in \Gamma_x$ and $l$ does not cross $\tilde{\Gamma}_y$.
\begin{subequations}
\begin{align}
& Z_l \leftrightarrow (b_{v_1} + b_{v_1}^\dagger) (b_{v_2} + b_{v_2}^\dagger), \\   
& X_l Z_{\SW(l)} \leftrightarrow \left[ \prod_{l \in \Gamma_{1,v}} (-1)^{b_v^\dagger b_v} \right] \
i \gamma_{p_1} \gamma'_{p_2} \ Z^W_2.
\end{align}
\end{subequations}

\item $l \notin \Gamma_x$ and $l$ crosses $\tilde{\Gamma}_y$.
\begin{subequations}
\begin{align}
& Z_l \leftrightarrow (b_{v_1} + b_{v_1}^\dagger) (b_{v_2} + b_{v_2}^\dagger) \
\left( \prod_{l \times \tilde{\Gamma}_{1,p}} -i \gamma_p \gamma'_p \right) \ X^W_2, \\
& X_l Z_{\SW(l)} \leftrightarrow i \gamma_{p_1} \gamma'_{p_2}.
\end{align}
\end{subequations}

\item $l \in \Gamma_x$ and $l$ crosses $\tilde{\Gamma}_y$.
\begin{subequations}
\begin{align}
& Z_l \leftrightarrow (b_{v_1} + b_{v_1}^\dagger) (b_{v_2} + b_{v_2}^\dagger) \ X^W_2, \\
& X_l \leftrightarrow i \gamma_{p_1} \gamma'_{p_2} \ Z^W_2.
\end{align}
\end{subequations}
\end{enumerate}
Here for a horizontal (vertical) link $l$, $v_1$ and $v_2$ are the two vertices
connected by it, $p_1$ and $p_2$ are the plaquettes to its top (left) and bottom (right).
Again, the non-local boson and fermion parities in the dual operators reflect the semionic
statistical interaction between $e$ and $\varepsilon$ particles. 
Moreover, when an $e$ ($\varepsilon$) particle is moving across the $x/y$-direction boundary,
there will be an associated $X_{2/1}^W$ ($Z_{1/2}^W$) operator.
The spin $Z$ configuration of WLS determines the boundary condition of $\varepsilon$ particles.

\section{Model Hamiltonian} \label{sec:model}
In this study, we consider Hamiltonians of the form:
$H = H_0 + H_1$.
$H_0$ commutes with $A_v B_{\mathsf{p}(v)}$ for $\forall v$, according to
the boson-fermion mapping introduced in \Secref{sec:boson-fermion mapping},
its dual operator has dynamical ($\varepsilon$) fermions and static $\pi$-fluxes ($e$ particles).
Many exactly solvable models can be constructed from these type of Hamiltonians by making the fermions free, e.g.,
the Kitaev honeycomb model~\cite{Kitaev2006,Chen2018, Rao2021,Pozo2021}.
$H_1$ will be a term that allows the motion of $e$ particles, while preserving their total number.
We choose $H_0$ to be given by:
\begin{widetext}
\begin{align} \label{eq:H_0}
H_0 = & \sum_{l \in h\text{-link}} -J_x \ X_l Z_{\SW(l)} + J_y \ Y_l Y_{\SE(l)}
-J_z \ X_{\NE(l)} Z_{l}
+ \kappa \ [ Z_l Z_{\SW(l)} Y_{\SE(l)}
+ X_l X_{\NE(l)} Y_{\SE(l)} - Y_{l}Z_{\SE(l)} X_{\NE(l)} ] \nonumber \\
& + \sum_{l \in v\text{-link}} \kappa \ [ Y_l Z_{\SW(l)} X_{\NE(l)} 
- Z_l Z_{\SW(l)} Y_{\NW(l)} - X_l X_{\NE(l)} Y_{\NW(l)} ].
\end{align}
\end{widetext}
Here $h/v$-link stands for horizontal/vertical links.
This Hamiltonian is equivalent to the Kitaev homeycomb Hamiltonian
(with a Haldane mass term $\kappa$)~\cite{Kitaev2006} by plaing the
sites of the original honeycomb lattice onto the links of a square lattice
(see Supplementary Section S-\rom{3}~\cite{SI} for a schematic of the lattice)
and a unitary transformation $U$ which transforms:
\begin{align}
X_j \leftrightarrow Z_j,\ Y_j \rightarrow -Y_j,\ \forall j \in A \text{-sublattice}.
\end{align}
Under the duality transformation introduced in \Secref{sec:boson-fermion mapping},
the dual Hamiltonian reads (for an infinite system):
\begin{widetext}
\begin{align}
\tilde{H}_0 = & -\sum_p \left(
J_x e^{i \pi \beta_{\tilde{\mathbb{L}}(p,p+\hat{y})}} i\gamma_{p+\hat{y}}\gamma'_p
+ J_y e^{i \pi \beta_{\tilde{\mathbb{L}}(p,p+\hat{y})}} i \gamma_{p+\hat{y}} \gamma'_{p+\hat{x}}
+ J_z i \gamma_p \gamma'_{p+\hat{x}} \right) \nonumber \\
& - \kappa \sum_p \left[
e^{ i\pi \beta_{\tilde{\mathbb{L}}(p,p+\hat{y}) } } i\gamma_{p+\hat{y}} \gamma_{p-\hat{x}}
+ e^{i \pi b_{\mathsf{v}(p)}^\dagger b_{\mathsf{v}(p)} } i \gamma_{p-\hat{x}} \gamma_p
+ e^{i \pi \beta_{\tilde{\mathbb{L}}(p,p+\hat{y})}} i \gamma_p \gamma_{p+\hat{y}}
\right. \nonumber \\
& \left. + e^{ i\pi \beta_{\tilde{\mathbb{L}}(p-\hat{x},p-\hat{x}+\hat{y})} } i\gamma'_{p+\hat{y}}\gamma'_p
+ i\gamma'_{p+\hat{x}} \gamma'_p + e^{i\pi \beta_{\tilde{\mathbb{L}}(p,p-\hat{y})} }
i\gamma'_{p-\hat{y}} \gamma'_{p+\hat{x}} \right].
\end{align}
\end{widetext}
Here $\tilde{\mathbb{L}}(p,p')$ stands for the link sandwiched by plaquettes $p$ and $p'$,
$\beta_l = \sum_{v \in L(l)} b_v^\dagger b_v$ for a horizontal link $l$
(here $L(l)$ stands for the vertices to the left of link $l$).
It is clear that $\tilde{H}_0$ has a Bogoliubov-de Gennes (BdG) form 
for the $\varepsilon$ fermions in a background of static $\pi$-fluxes ($e$ particles),
and can be solved exactly within any given real-space configuration of $e$ particles.

As for $H_1$, we choose it to be:
\begin{align} \label{eq:H_1}
H_1 = g \sum_l Z_l \frac{ 1 - A_{v_1(l)} B_{\mathsf{p}(v_1(l))}
A_{v_2(l)} B_{\mathsf{p}(v_2(l))} }{2},
\end{align}
with $v_1(l)$ and $v_2(l)$ being the two vertices adjacent to link $l$.
Its dual operator $\tilde{H}_1$,
according to \cref{eq:esp-e,eq:bf-mapping-Z_l}, reads (for the infinite lattice case):
\begin{align}
\tilde{H}_1 = & g \sum_v
b_v^\dagger b_{v+\hat{y}} \prod_{p \in R( \mathbb{L} (v,v+\hat{y}) )}
\left( -i \gamma_p \gamma'_p \right) \nonumber \\
& + b_v^\dagger b_{v+\hat{x}} + h.c.
\end{align}
$\sim b_{v_1}^\dagger b_{v_2} + b_{v_2}^\dagger b_{v_1}$,
Here $\mathbb{L}(v,v')$ stands for the link connecting vertices $v$ and $v'$,
$R(l)$ stands for the plaquettes to the link $l$.
Notice that the above Hamiltonian is a sum of operators that act on spins
contained within some local region of the link $l$, and therefore it is a
strictly local perturbation
(even though it contains products of several spin operators).
So $\tilde{H}_1$ contains nearest-neighbor $e$-particle hopping terms.
Note that it is also dressed by $\varepsilon$ particles' parities due to the
statistical interaction between $e$ and $\varepsilon$ particles.
To perform calculations, in this study, $H_1$ will be treated as
a perturbation to $H_0$.

\section{Berry phases of visons} \label{sec:BP-visons}
\subsection{Kitaev model with a Haldane mass term} \label{sec:Kitaev-Haldane}
We will use the previously described mapping to compute the Berry phase for transporting
the $\pi$-flux in a closed loop around a single plaquette. This phase can be viewed as a universal
characterization of the topologically ordered state enriched by lattice translational symmetry~\cite{Wen2002,Essin2013,Barkeshli2019,Lu2016,Wen2003, 
Kou2008,Mesaros2013}. 

\begin{figure}[t]
\centering
\includegraphics[width=0.49\textwidth]{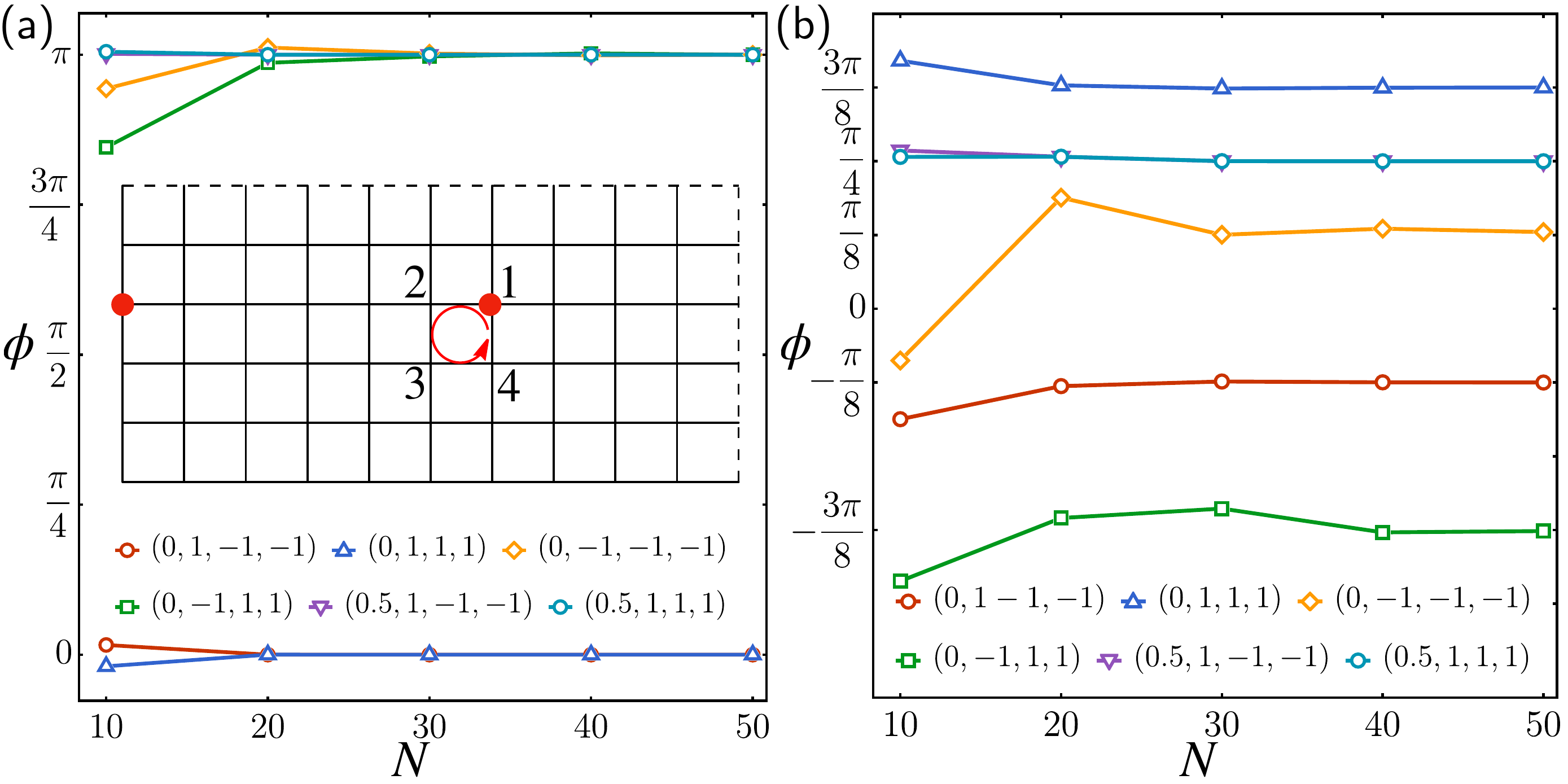}
\caption{
Berry phase for $e$ particles.
(a) Berry phase for a single-plaquette movement of an $e$-particle.
$\phi \rightarrow 0$ or $\pi$ in the thermodynamic limit.
The inset indicates the set-up of numerical calculations:
two $e$ particles are highlighted by the red dots with one of them
hops circularly between the 4 sites. The legends indicate values
of parameters $(t,J_y,z_1,z_2)$ in the model.
(b) Berry phase for the exchange of two $e$ particles.
$\phi$ converges to predicted values in
Ref.\cite{Kitaev2006} as $N \rightarrow \infty$.
}
\label{fig:BP}
\end{figure}

In order to compute the Berry phase, we place two $e$ particles far apart
on a torus, and will allow only one of them to move within the 4 vertices surrounding a
plaquette (see inset \Fref{fig:BP}(a)). This is accomplished by only adding the flux-hopping operator,
from \Eqref{eq:H_1}, to be non-zero at the links connecting these 4 vertices.
For a fixed WLS configuration $| z_1, z_2 \rangle$,
when the mobile $e$-particle is located at site $j \in \{1,2,3,4\}$,
the corresponding \emph{physical} (even number of $\varepsilon$ particles)
ground state  of $\tilde{H}_0$ reads:
\begin{equation}
| \Phi_j \rangle = b_0^\dagger b_j^\dagger|0\rangle \otimes
|\Psi_j^\varepsilon \rangle \otimes |z_1,z_2 \rangle.
\end{equation}
Here $| \Psi_j^\varepsilon \rangle$ is the even-parity ground state of
a BdG Hamiltonian with two $\pi$ fluxes at $0$ and $j$, and the
$z_{1,2} \teq \pm 1$ are the eigenvalues of the Wilson loop operators that label the global
periodic/antiperiodic boundary conditions of the fermions along the $x$- and $y$-directions
(see Supplementary Section S-\rom{3}~\cite{SI}).
$| \Psi_j^\varepsilon \rangle$ can be solved exactly and has a BCS form
(see Supplementary~\cite{SI}).
The Berry phase for this close-loop movement of an $e$-particle is:
$e^{i \phi} \approx \prod_{j=1}^4 \langle \Phi_{j+1} | Z_{j+1,j} | \Phi_j \rangle$.
Note that the index $j$ runs cyclically from 1 to 4,
i.e., $| \Phi_{5} \rangle \equiv | \Phi_1 \rangle$.
In the dual space, the Berry phase reads:
\begin{align}
e^{i \phi} = &
\langle \Psi_1^\varepsilon | \left( \prod_{p\in L(4,1)} -i \gamma_p \gamma'_p \right)
| \Psi_4^\varepsilon \rangle
\langle \Psi_4^\varepsilon| \Psi_3^\varepsilon \rangle \nonumber \\
& \times \langle \Psi_3^\varepsilon |
\left( \prod_{p \in L(3,2)} -i \gamma_p \gamma'_p \right)
| \Psi_2^\varepsilon \rangle
\langle \Psi_2^\varepsilon | \Psi_1^\varepsilon \rangle.
\end{align}
Here $L(4,1)$ denotes the string of plaquettes to the left of link $(4,1)$
that runs until the left edge of the torus.

In our study, we take the following parameters:
$J_x \teq J_z \teq 1, \kappa \teq 0.1$.
The torus has $N \times N$ plaquettes with $N$ even.
We consider two values $J_y \teq \pm 1$ which corresponds to fermionic BdG states 
with Chern numbers $C \teq \pm 1$.
There are 4 high-symmetry points (HSPs) in $k$-space which are unpaired in a
BdG Hamiltonian~\cite{Rao2021,Kou2009,Kou2010}:
$(0, 0)$, $(\pi, 0)$, $(0, \pi)$ and $(\pi, \pi)$.
For $J_y \teq 1$, the fermion band energy $\epsilon(0,0) \tlt 0$ and is positive at 
other three HSPs. In this case, we have found that the single-plaquette Berry phase
$\phi \rightarrow 0$ with increasing $N$ for any BC.
On the other hand, for $J_y \teq -1$, $\epsilon(\bk) \tlt 0$ at $(0,0)$, $(\pi, 0)$
and $(0, \pi)$, and is positive at $(\pi, \pi)$. For this case we have found that
for any BC, $\phi \rightarrow \pi$ as $N$ increases.
The results are presented in \Fref{fig:BP}(a) and this is one of the main findings of
our study. 

The motion of the vison in the ferromagnetic (FM) and antiferromagnetic (AFM) Kitaev
models induced by physically realistic perturbations such as the Zeemann field,
has also been studied in Refs.~\cite{Joy2022,Chen2022a}.
While an earlier version of Ref.~\cite{Joy2022} had concluded that the phase of vison
in the FM model was $\pi$ around a unit cell, the updated understanding provided
in Refs.~\cite{Joy2022,Chen2022a} is currently in mutual agreement with the conclusion
that the vison acquires zero phase in the FM model and $\pi$ phase in the AFM model
around a unit cell, which is also in agreement with the current study.


We also studied the braiding phases for two anyons.
To avoid geometric phases depending on the details of the braiding path,
we follow the Levin-Wen protocol \cite{Levin2003,Kitaev2006,Kawagoe2020}.
\Fref{fig:BP}(b) presents results of the braiding phases. For $J_y \teq 1$,
with increasing system size, the braiding phase $\phi \rightarrow -\pi/8$ for
anti-periodic BC (APBC) and $\phi \rightarrow 3\pi/8$ for periodic BC (PBC).
While for $J_y \teq -1$, the $\phi \rightarrow \pi/8$ for APBC and
$\phi \rightarrow -3\pi/8$ for PBC.
Our results for $\phi$ match exactly the prediction of
$R_1^{\sigma,\sigma} \propto \exp{(-i C \pi/8)}$ and
$R_\varepsilon^{\sigma,\sigma} \propto \exp{(i C 3\pi /8)}$ in Ref.~\cite{Kitaev2006}
(here $\sigma$ stands for the $\pi$-flux particle).
The difference between PBC and APBC originates from the fermion ground state parity
of $\tilde{H}_0$.
The state with $J_y \teq 1$ is a $p + i p$ topological superconductor,
and the ground state would have an odd number of fermions under PBC~\cite{Read2000},
which is unphysical in our case.
Since, only even-parity states are physical, the lowest energy
physical eigenstate of $\tilde{H}_0$ in this case is actually the first excited state of
the BdG Hamiltonian with a single Bogoliubov quasiparticle. Thus 
for PBCs the $\pi$-fluxes are in the fusion sector
$\sigma \times \sigma \teq \varepsilon$, explaining the difference in braidings that we
observe in \Fref{fig:BP}.
As for APBC, the ground state of the BdG Hamiltonian contains an even number
of fermions, therefore the $\pi$-fluxes are in the fusion sector
$\sigma \times \sigma \teq 1$.

\subsection{Higher Chern numbers and conjecture for arbitrary case} \label{sec:higher-Chern}
One can also explore cases with higher Chern numbers by correspondingly modifying
$H_0$. This illustrates the power of this construction allowing to write an exactly
solvable model for any free fermion Hamiltonian of interest.
We accomplished this explicitly by introducing some $4$-spin interaction terms to
$H_0$ in \Eqref{eq:H_0}:
\begin{align}
 & \frac{t}{2} \left[
 \sum_{l \in h\text{-link}} ( Y_l Z_{\SW(l)} Z_{\NE(l)} Y_{\N(l)}
 + Y_l Y_{\W(l)} X_{\SW(l)} X_{\NE(l)} ) \right. \nonumber \\
 & + \sum_{l \in v\text{-link}} ( Y_l Y_{\text{S}(l)} X_{\SW(l)} X_{\NE(l)}
 + Y_l Y_{\E(l)} Z_{\SW(l)} Z_{\NE(l)} ) \nonumber \\
 & \left. + \sum_p B_p + \sum_v A_v \right].
\end{align}
The $\E(l)$ ($\text{S}(l)$) stands for the link to the east (south) of $l$ within a
common plaquette.
Under the duality mapping established in this work, these new terms are mapped
to third-neighbor Majorana fermion couplings of the form:
\begin{align}
t \sum_p \left( -i \gamma_p \gamma'_p -i \gamma_p \gamma'_{p+2\hat{x}}
 - i \gamma_p \gamma'_{p-2\hat{y}} \right).
\end{align}
Here for simplicity we have omitted the non-local vison parities and the WLS
operators involved in some of the terms, for the complete expression see
Supplementary Section \rom{3}~\cite{SI}.

At $J_y \teq 1$, $t \teq 0.5$, $\tilde{H}_0$ has $C \teq -2$.
$\epsilon_k \tlt 0$ at all HSPs, so for both PBC and APBC,
the fermion ground state parity of $\tilde{H}_0$ is even.
There are two types of anyons in this case \cite{Kitaev2006} and we studied
the sector with $a \times \Bar{a} \teq 1$ where $a$ and $\bar{a}$ denote the two kinds of
$\pi$-flux particles in these states.
When braiding a single $e$-particle around a plaquette, we found Berry phase 
$\phi \teq \pi$ for any BC.
As for the braiding phase, we obtained $R_{1}^{a,\Bar{a}} \teq
e^{i \pi/4} \teq e^{-i C \pi/8 }$, which is also consistent with Ref.~\cite{Kitaev2006}.
More details can be found in Supplementary Information~\cite{SI}.

As mentioned before, the phase $\phi$ acquired by a $\pi$-flux upon enclosing a plaquette
is a universal characteristic of the symmetry enriched topological state.
BdG states of fermions with lattice translations can be classified by their Chern number,
$C \in \bbZ$, and 4 parity indices $\zeta_{k}$, which dictate whether the band is inverted 
($\zeta_k \teq -1$) or not ($\zeta_k \teq 1$) in each of the 4 HSPs of the Brillouin zone~\cite{
Rao2021,Kou2010,Kou2009,Sato2010,Sato2010L,Geier2020,Ono2020}.
Therefore the value of $\phi$ should be a function uniquely fixed by $C$ and $\zeta_k$.
The analytical proof of the value of $\phi$ in the most general case is not known to us.
Ref.~\cite{Rao2021} showed that when $C \teq 0$, $\phi \teq 0$ when all $\zeta_k \teq 1$ and
$\phi \teq \pi$ when all $\zeta_k \teq -1$ (all HSP are band-inverted),
in agreement with previous arguments~\cite{Essin2013}.
Ref.~\cite{Rao2021}  also showed that the cases with $C \teq 0$ and only two $\zeta_k \teq -1$,
corresponds to states with ``weak symmetry breaking'' (and thus the $\pi$-fluxes cannot be
transported to adjacent vertices with local operations).
We have shown here that when only one $\zeta_k \teq -1$ and $C \teq 1$ then $\phi \teq 0$,
and when three $\zeta_k \teq -1$ and $C \teq -1$ then $\phi \teq \pi$.
We also showed that when $C \teq -2$ and all four $\zeta_k \teq -1$, then $\phi \teq \pi$.
This suggest the conjecture that for states with odd $C$ and only one $\zeta_k \teq -1$,
then $\phi \teq 0$ and states with three $\zeta_k \teq -1$ then $\phi \teq \pi$.
For states with even $C$ and all $\zeta_k \teq 1$ then $\phi \teq 0$ and those with all four
$\zeta_k \teq -1$ then $\phi \teq \pi$ (states with even $C$ and only two $\zeta_k \teq -1$ should
display weak symmetry breaking of translations~\cite{Rao2021}).

\section{Discussions} \label{sec:discuss}
We have established an \emph{exact} mapping between a 2D spin system and
a 2D boson-boson $(e,m)$ or boson-fermion $(e, \varepsilon)$ system,
where the two types of particles in the dual space are mutual semions,
which generalizes the previous dual maps that relied on imposing local $\bbZ_2$ 
constraints~\cite{Chen2018}.
This amounts to constructing explicit vison and spinon non-local
creation/annihilation operators in terms of the underlying spin degrees of freedom.
Based on this mapping, we found that the Berry phase for the transport of the vison
($\pi$-flux excitation) around a single plaquette was
quantized to be $0$ or $\pi$.
We have conjectured a universal form of this phase that depends on the Chern number
and the parity indices at HSPs of the BdG band structure of the spinons,
generalizing previous results from non-chiral states in Refs.~\cite{Essin2013,Rao2021}
to chiral and non-abelian states.
We also computed explicitly the braiding phase between two visons,
which was found to be consistent with the general arguments of Ref.~\cite{Kitaev2006}
for both $C \teq 1$ and $C \teq 2$ states of the spinons.
In the models studied here, the $e$-particles are static and we only need to
solve a free fermionic Hamiltonian of $N^2 \times N^2$. Thus the Berry phase 
for $e$-particle movement can be calculated even for relatively large system sizes
without too much computational cost.
The lattice dualities developed in this work are universal and can be used to study
not only the Berry phases of translations of visons but many other topological and
dynamical properties of these excitations, such as their effective mass and dispersions,
which can be crucial in understanding their role in real materials and
experiments~\cite{Joy2022}.

\begin{acknowledgments}
C.C. thanks Guo-Yi Zhu for enlightening discussions.
C.C. and P.R. thank Professor Peter Fulde for kind encouragement on persuing this
project.
C.C. acknowledges the support from Shuimu Tsinghua Scholar Program.
\end{acknowledgments}

\appendix

\bibliography{reference.bib}



\clearpage
 \includepdf[pages={{},1,{},2,{},3,{},4,{},5,{},6,{},7,{},8,{},9,{},10,{},11,
 {},12,{},13,{},14,{},15,{},16}]%
 {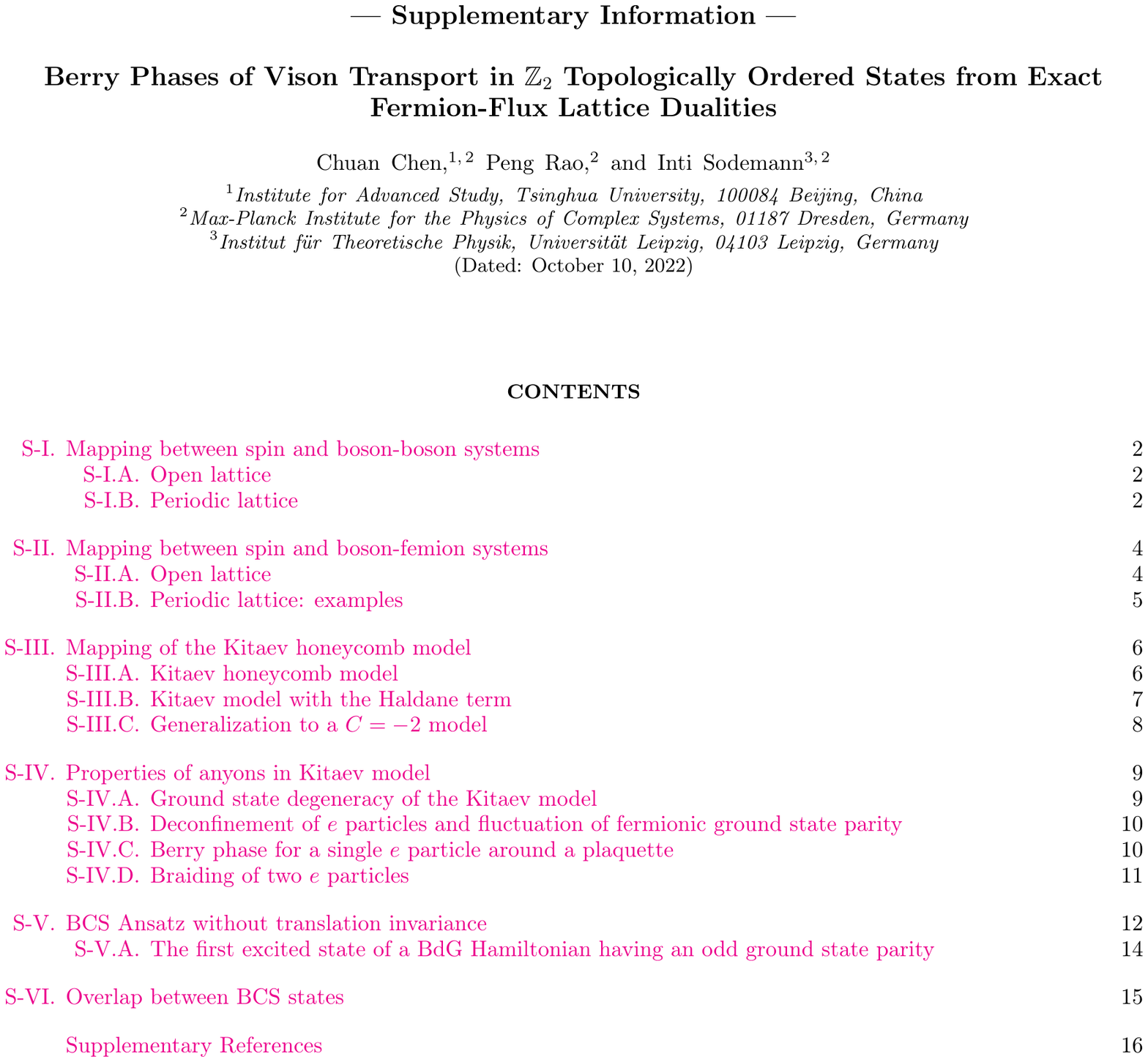}

\end{document}